\pdfoutput=1
\RequirePackage{atlaslatexpath}

\documentclass[cernpreprint, atlasdraft=true, texlive=2016, coverpage=false, UKenglish, ]{atlasdoc}
\usepackage[backend=bibtex]{atlaspackage}
\usepackage{atlasbiblatex}
\usepackage{atlasphysics}
\usepackage{comment}

\graphicspath{{logos/}{figures/}}

\AtlasTitle{Study of  the KNO scaling  in   \( pp\)  collisions  at   \(\sqrt{s}\) from \(0.9\) to \(13\)~TeV 
using   results of the ATLAS  at the LHC
\\[10mm]
\centerline{\LARGE Yuri~Kulchitsky, Pavel~Tsiareshka}
\centerline{\Large Joint Institute for Nuclear Research}
}
 
\AtlasAbstract{
The comparisons of the charged-particle multiplicity and  the average transverse momentum distributions 
on the scaled multiplicity, KNO scale, using the results of the ATLAS collaboration at the LHC 
are presented.
These distributions 
were measured 
in proton-proton collisions at centre-of-mass energies from 0.9 to 13~TeV 
for the absolute pseudorapidity region less than 2.5 and  two samples of events with 
each 
charged-particle transverse 
momentum greater than  100 and 500~MeV, respectively. 
The shape evolution of the multiplicity distributions with a collision energy 
is studied 
in terms of KNO scaling variables.
The charged-particle multiplicity distributions on the KNO scale have the 
similar
shape and decrease with increasing collision energy. 
The KNO  distributions tend 
to be 
independent of energy for the highest energies.
The average transverse momentum distributions on the KNO scale 
have 
a similar shape and increase with increasing collision energy. 
}

 \AtlasDate{\today}

 \arXivId{2202.06697}
 
\PreprintIdNumber{arXiv:2202.06697 [hep-ex]}

\AtlasJournal{EPJC} 

\begin{document}

\maketitle
   


\section{Introduction}
\label{intro}  

%
%

The study of charged-particle distributions in proton-proton (\(pp\)) collisions  
probes the strong interaction in the low-momentum transfer, non-perturbative region of 
quantum chromodynamics.

Charged-particle distributions  were measured  at 
CERN's  Large Hadron Collider (LHC)  \cite{Evans:2008zzb} experiments 
for various  centre-of-mass energies,  \( \sqrt{s} \),   from 0.9 to 13~TeV   by 
the ATLAS \cite{PERF-2007-01}  Collaboration 
\cite{STDM-2010-01,STDM-2010-06,STDM-2014-19,STDM-2015-02,STDM-2015-17}, 
the CMS \cite{CMS:2008xjf}  Collaboration 
\cite{CMS:2010wcx,CMS:2010tjh,CMS:2010qvf,CMS:2015zrm,CMS:2018nhd}, 
the CMS and TOTEM  \cite{TOTEM:2008lue} Collaborations 
\cite{CMS:2014kix}, 
the TOTEM  Collaboration
\cite{TOTEM:2014zyx}, 
the ALICE  \cite{Crochet:2008zz} Collaboration
\cite{ALICE:2010cin,ALICE:2010mty,ALICE:2015qqj,ALICE:2015olq,ALICE:2017pcy,ALICE:2019dfi}
and the LHCb  \cite{LHCb:2008vvz} Collaboration
\cite{LHCb:2011jir,LHCb:2014wmv}. 
Charged-particle distributions  were studied by the CDF Collaboration at Tevatron (Fermilab) 
at  \( \sqrt{s} = 0.63\), \(1.8\)  and \(1.96\)~TeV  
\cite{CDF:1989nkn,CDF:2009cxa} 
and 
by the UA1, UA4 and UA5 Collaborations at the SPS  (CERN) 
at  \( \sqrt{s} = 0.2\), \(0.54\) and  \(0.9\)~TeV
\cite{UA4:1985idi,UA5:1985hzd,UA5:1988gup,UA1:1989bou}. 
 
Measurements of charged-particle distributions in the ATLAS experiment
\cite{STDM-2010-01,STDM-2010-06,STDM-2014-19,STDM-2015-02,STDM-2015-17}
at centre-of-mass energies \(\sqrt{s}=0.9\),  \(2.36\), \(7\),  \(8\) and \(13\)~TeV  
were performed for  the  pseudorapidity region \(\mid\eta\mid <2.5\) 
and  for two samples of events:    with primary charged-particle multiplicity, \(n_{\mathrm{ch}}\),
more than  or equal to 2 and 1  and with  the 
each charged-particle  transverse momentum 
\(p_{\mathrm{T}}\)  more  than \(100\) and \(500\)~MeV,  respectively. 

The hypothesis that at very high   energies  the probability distributions  \( P (n, \sqrt{s}) \) 
of producing \( n \)  particles in a certain collision process should exhibit 
a scaling relation was  proposed  in  Refs.~\cite{Polyakov:1970lyy,Koba:1972ng,Proceedings:1973mza}.
%
This scaling behaviour is a property of particle multiplicity distributions known as the KNO scaling hypothesis.
The  main assumption of KNO scaling is Feynman scaling   \cite{Feynman:1969ej},
where was  concluded that for asymptotically large energies with  \(\sqrt{s} \rightarrow  \infty \) 
the mean total number of any kind of particle rises logarithmically 
with a centre-of-mass energy as  
\(  \langle n \rangle  \propto \ln{\sqrt{s}}\).
For this assumption the multiplicity distribution  \( P (n, \sqrt{s}) \) 
was represented as
\begin{equation}
\label{eq_pn2}
P (n, \sqrt{s}) =  \frac{1}{\langle n (\sqrt{s}) \rangle} \Psi ( z )  + 
\mathcal{O} \left( \frac{1}{\langle n (\sqrt{s}) \rangle^{2}} \right) ,
\end{equation}
where \( \langle n (\sqrt{s}) \rangle \)  is the average multiplicity of primary particles at centre-of-mass energy,  
\( \Psi ( z ) \) is a particle distribution as a function of the scaled multiplicity
\( z =  { n (\sqrt{s}) }/{ \langle n (\sqrt{s}) \rangle } \). 
%
The first term in Eq.\ (\ref{eq_pn2}) was results from the leading term in \( \ln{\sqrt{s}}  \)
(KNO scaling hypothesis)  
and the second term contains all other terms
\cite{Grosse-Oetringhaus:2009eis}. 
The multiplicity distributions become simple rescaled copies of the universal function  \( \Psi ( z ) \) 
depending only on the scaled multiplicity or an energy-independent function.
Asymptotically for 
\(\sqrt{s} \rightarrow  \infty \)  
the second term in Eq.~(\ref{eq_pn2}) is tends to zero and therefore  KNO scaling holds. 

The energy independence of the moments 
\( C_{\mathrm{q}} (\sqrt{s}) = 
{\langle n^q (\sqrt{s}) \rangle }/{\langle n (\sqrt{s}) \rangle^q} \)
in an energy asymptotic,  \(\sqrt{s} \rightarrow  \infty \),  
was the precise finding of the KNO scaling  \cite{Koba:1972ng}.
The introduction of the novel physically well-motivated scaling rules for high-energy data 
was presented in Ref.~\cite{Hegyi:2000sp}.

The KNO scaling was studied at the LHC energies by  the CMS \cite{CMS:2010qvf} 
and the ALICE  \cite{ALICE:2010cin,ALICE:2015olq} Collaborations.
The KNO scaling violation  were  observed for a larger rapidity range in  LHC experiments at 
centre-of-mass energies  \(\sqrt{s} = 0.9\) -- \(8\)~TeV \cite{CMS:2010qvf,ALICE:2010cin,ALICE:2015olq}.  

Charged-particle multiplicity and transverse momentum distributions in \(pp\) collisions 
at  centre-of-mass energies    \(\sqrt{s} = 0.2\) -- \(14\)~TeV   
within the Monte Carlo Quark-Gluon String Model (MC QGSM) 
\cite{Kaidalov:1982xg,Kaidalov:1982xe}  
based on Gribov’s Reggeon field theory  \cite{Gribov:1967vfb,Gribov:1983ivg}
were  studied  in  Refs.~\cite{Bleibel:2010ar,Bravina:2016cme}, 
where special attention was given to the origin of violation of the KNO scaling.
Detailed theoretical description of the KNO scaling was done in 
Refs.~\cite{Kittel:2005fu,Dremin:2000ep,Grosse-Oetringhaus:2009eis}.

This publication presents 
in Sec.~\ref{Charged-particle} the comparison of  the charged-particle distributions as a function on the KNO scale, 
in Sec.~\ref{KNO_scaling} the study of the KNO scaling and
in Sec.~\ref{average_pT_z} the comparison of  the average transverse momentum of the primary charged particles   
as a function on the KNO scale  based on the results of the ATLAS Collaboration.
The moments \( C_{\mathrm{q}} \) were not studied by the ATLAS Collaboration 
and is not discussed in the paper.

\section{Charged-particle multiplicity distributions}
\label{Charged-particle}

The measurements used for the analysis are the data  on \(pp\) collisions at  \( \sqrt{s} = 0.9\) -- \(13\)~TeV 
recorded by the  ATLAS experiment \cite{PERF-2007-01}   at the LHC \cite{Evans:2008zzb} in 2010 -- 2015
\cite{STDM-2010-01,STDM-2010-06,STDM-2014-19,STDM-2015-02,STDM-2015-17}. 
The data were taken in the special  configuration of the LHC with low beam currents and reduced beam focusing, 
producing a low mean number of interactions per bunch-crossing in the range \(0.003\) -- \(0.007\).

The following measured  observables  are used in the analysis:
\( P (n_{\mathrm{ch}}, \sqrt{s})  =
({d N_{\mathrm{ev}} (\sqrt{s})}/{ d n_{\mathrm{ch}}})/{ N_{\mathrm{ev}} (\sqrt{s}) } \)
and
\(  \langle p_{\mathrm{T}}  ( n_{\mathrm{ch}}, \sqrt{s}) \rangle ,  \)
where 
\(N_{\mathrm{ev}} \) 
is the 
number of events with  primary charged particles in the kinematic acceptance, 
\(n_{\mathrm{ch}}\) 
is the number of primary charged particles within the kinematic acceptance in an event
and 
\(\langle  p_{\mathrm{T}} ( n_{\mathrm{ch}}, \sqrt{s}) \rangle\) 
is the average momentum component transverse  to the beam direction for charged particles in an event
at  centre-of-mass  energy \( \sqrt{s} \). 

As was defined in Sec.~\ref{intro},   for the verification of  the KNO scaling hypothesis  the following  
equation with  dependence from  a centre-of-mass energy, \(\sqrt{s}\),  and a kinematic region, 
\( p_{\mathrm{T}}^{\mathrm{min}} \),  was used:
\begin{equation}
\label{eq_Psi_nch}
\Psi ( z , \sqrt{s})  
= { \langle n_{\mathrm{ch}} (\sqrt{s}, p_{\mathrm{T}}^{\mathrm{min}}) \rangle }  
\cdot  
P (n_{\mathrm{ch}}, \sqrt{s}, p_{\mathrm{T}}^{\mathrm{min}})  
= \frac{ \langle n_{\mathrm{ch}} (\sqrt{s}, p_{\mathrm{T}}^{\mathrm{min}}) \rangle }{ N_{\mathrm{ev}} (\sqrt{s},  p_{\mathrm{T}}^{\mathrm{min}} ) } \cdot \frac{d N_{\mathrm{ev}} (\sqrt{s}, p_{\mathrm{T}}^{\mathrm{min}} )}{ d n_{\mathrm{ch}}}.
\end{equation}

For correct comparison of  charged-particle multiplicity and average transverse momentum distributions 
for different energies or kinematic regions the scaled multiplicity is introduced as follows:
\begin{equation}
\label{eq_mch}
z =
\frac{ n_{\mathrm{ch}} (\sqrt{s}, p_{\mathrm{T}}^{\mathrm{min}}) }{
\langle 
n_{\mathrm{ch}} (\sqrt{s}, p_{\mathrm{T}}^{\mathrm{min}}) 
\rangle }. 
\end{equation}
For example,  comparison of results for different kinematic regions,  
with two \( p_{\mathrm{T}}^{\mathrm{min}} \) thresholds,  
was presented in Ref.~\cite{ATLAS:2022wvk}.

\begin{figure*}[h!]
\begin{minipage}[h]{0.50\textwidth} 
\center{\includegraphics[width=1.0\linewidth]{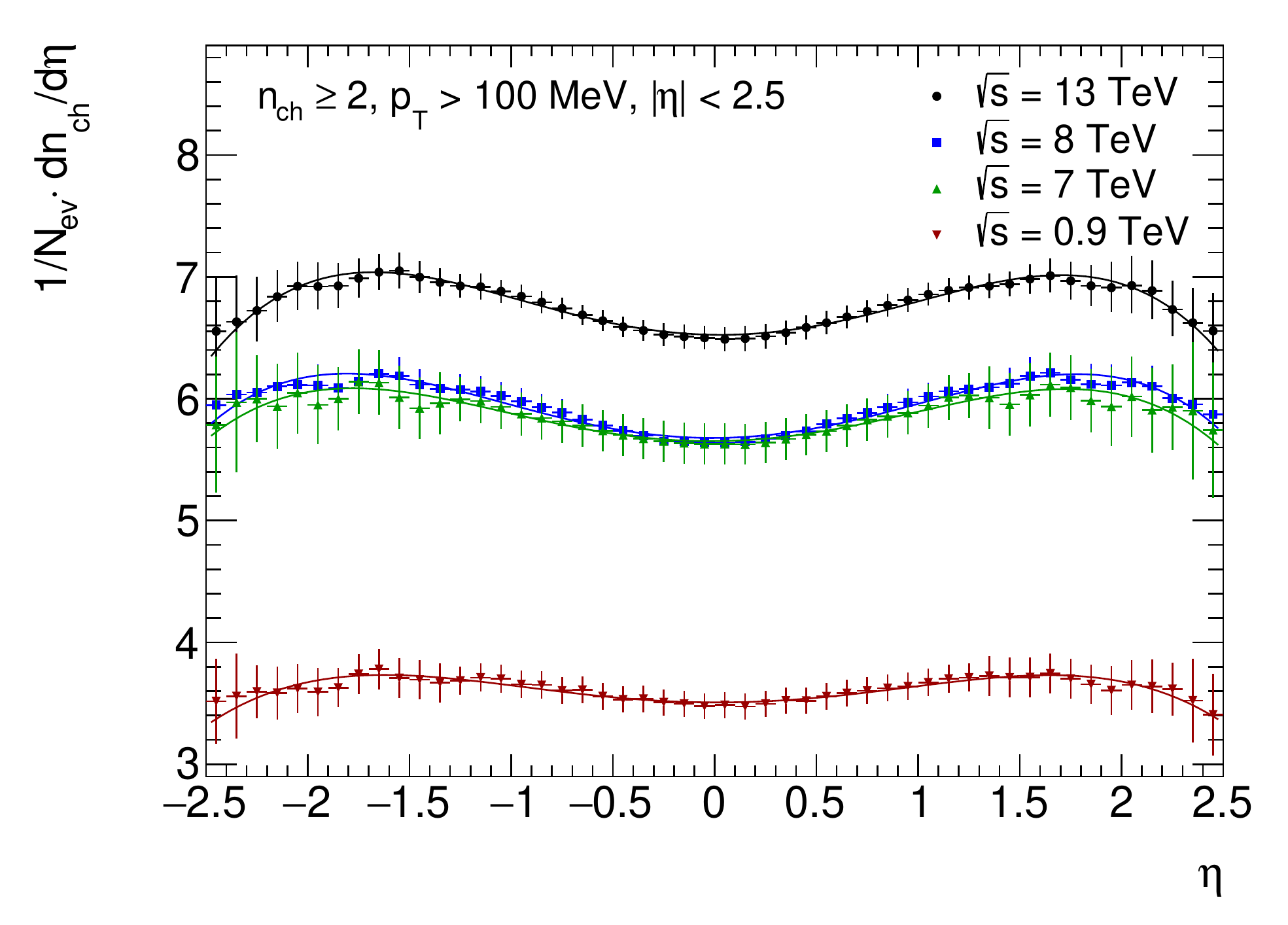}}  
(a)
\\
\end{minipage}
\hfill
\begin{minipage}[h]{0.50\textwidth} 
\center{\includegraphics[width=1.0\linewidth]{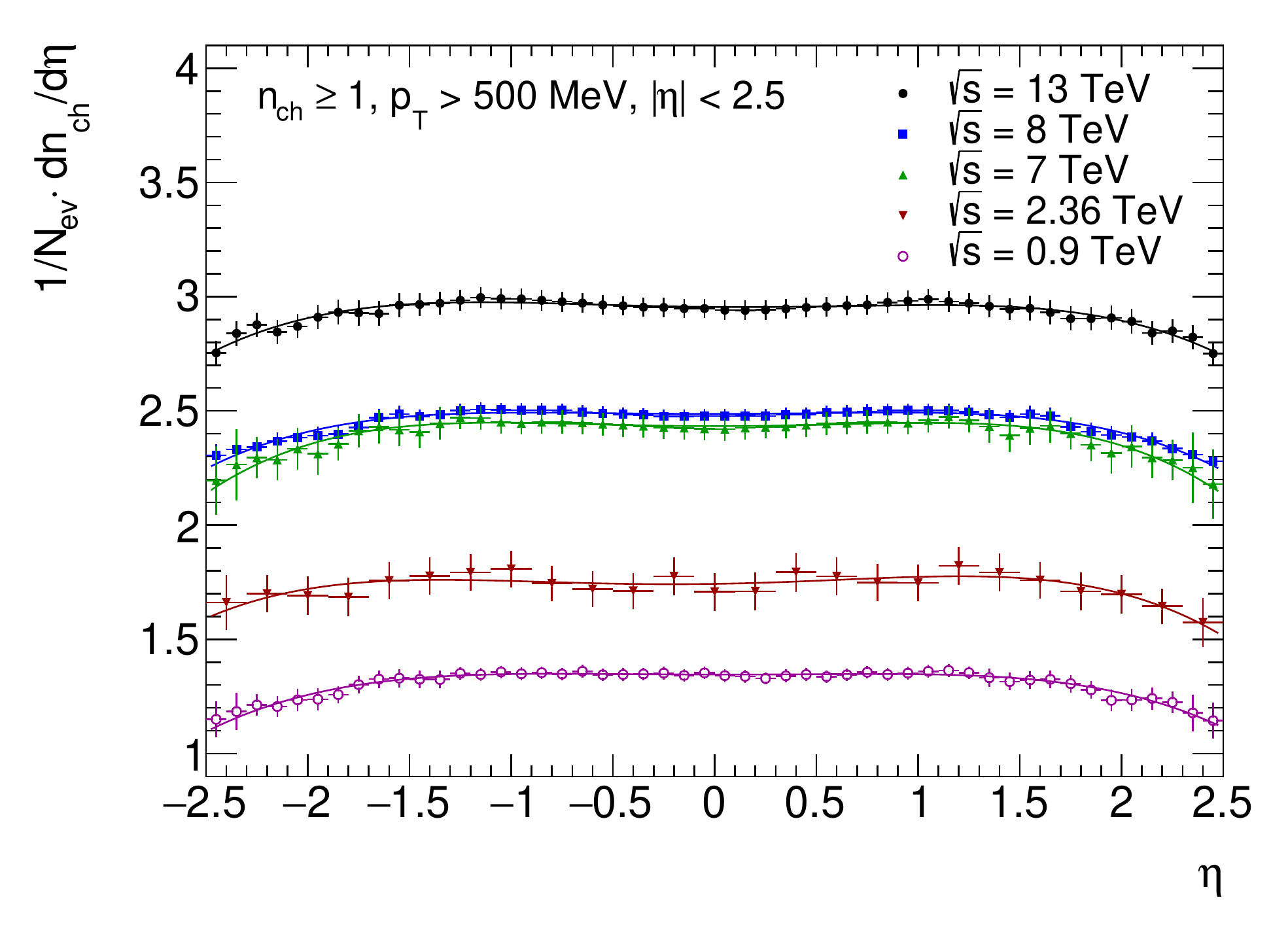}}  
(b)
\\
\end{minipage}
\caption{
The primary charged-particle average multiplicity,  
\( 1/N_{\mathrm{ev}} \cdot d n_{\mathrm{cn}}/d\eta \),  
dependence on pseudorapidity region \(-2.5 < \eta < 2.5\)  for the ATLAS Collaboration 
results  for the charged-particle  with  
(a) \(n_{\mathrm{ch}} \ge 2\), \(p_{\mathrm{T}} > 100\)~MeV  
and 
(b)  \(n_{\mathrm{ch}} \ge 1\), \(p_{\mathrm{T}} > 500\)~MeV 
at  centre-of-mass energies 
\(\sqrt{s} = 0.9\), \(2.36\), \(7\), \(8\) and \(13\)~TeV
\cite{STDM-2010-01,STDM-2010-06,STDM-2014-19,STDM-2015-02,STDM-2015-17}.
The coloured symbols represent the data. 
%
The vertical bars represent statistical and systematic uncertainties added in quadrature.
The black curves show the results of the fits with the fourth-degree polynomial function. 
}
\label{fig_events_average_eta}
\end{figure*}

\begin{table*}[t!]
\centering 
\caption{
The   average multiplicity, 
\( \langle n_{\mathrm{ch}} (\sqrt{s}, p_{\mathrm{T}}^{\mathrm{min}})\rangle  \),
as the  results of  the  fits with a polynomial function of the primary charged-particle average  multiplicity distributions 
on pseudorapidity region \(-2.5 < \eta < 2.5\)  
for the events samples with  \(p_{\mathrm{T}} > 100\)~MeV   and \(p_{\mathrm{T}} > 500\)~MeV 
at  centre-of-mass energies \(\sqrt{s} = 0.9\), \(2.36\), \(7\), \(8\) and \(13\)~TeV
using the ATLAS Collaboration results
\cite{STDM-2010-01,STDM-2010-06,STDM-2014-19,STDM-2015-02,STDM-2015-17}.
%
}
\label{tab:average_nch}
\medskip
 \begin{tabular}{ccrr}
\hline
\hline

\(\sqrt{s}\) 
& 
\(p_{\mathrm{T}}^{\mathrm{min}}\) 
& 
Average 
& 
Relative 
\\

[TeV]
& 
[MeV]
& 
Multiplicity 
& 
Uncertainty
\\

& 
& 
\( 
\langle n_{\mathrm{ch}} (\sqrt{s}, p_{\mathrm{T}}^{\mathrm{min}})\rangle  
\)
& 
\( 
\frac{\delta \langle n_{\mathrm{ch}} (\sqrt{s}, p_{\mathrm{T}}^{\mathrm{min}})\rangle}{ \langle n_{\mathrm{ch}} (\sqrt{s}, p_{\mathrm{T}}^{\mathrm{min}})\rangle } 
\)
\\
\hline
13		& \( 100\) 	& 	33.88$\pm$0.11	& 0.0032	
		\\
\cline{2-4}
		& \( 500\) 	& 	14.66$\pm$0.04	& 0.0027	
		\\
\hline
8		& \( 100\) 	& 	29.81$\pm$0.10	& 0.0034
		\\
\cline{2-4}	
		& \( 500\)	& 	12.25$\pm$0.03	& 0.0024 	
		\\
\hline
7		& \( 100\) 	& 	29.40$\pm$0.19	& 0.0065 	
		\\
\cline{2-4}
		& \( 500\) 	& 	11.98$\pm$0.05	& 0.0042 	
		\\
\hline
2.36	& \( 500\) 	& 	8.66$\pm$0.51		& 0.0589 	
		\\
\hline
0.9		& \( 100\) 	& 	18.06$\pm$0.12	& 0.0066 	
		\\
\cline{2-4}
		& \( 500\) 	& 	6.53$\pm$0.03		& 0.0046 	
		\\
\hline
\hline
\end{tabular}
\end{table*}

A fit with a fourth-degree polynomial function of the  primary charged-particle average multiplicity distributions on 
pseudorapidity region \(-2.5 < \eta < 2.5\) was used for  the calculation of an average multiplicity, 
\( \langle n_{\mathrm{ch}} ( \sqrt{s},  p_{\mathrm{T}}^{\mathrm{min}} ) \rangle \),
for different  centre-of-mass energies and \(p_{\mathrm{T}}^{\mathrm{min}} \) using the ATLAS results \cite{STDM-2010-01,STDM-2010-06,STDM-2014-19,STDM-2015-02,STDM-2015-17}.
The \( 1/N_{\mathrm{ev}} \cdot d n_{\mathrm{cn}}/d\eta \) distributions on pseudorapidity
are shown  in Fig.~\ref{fig_events_average_eta}.
The average multiplicity, 
\( \langle n_{\mathrm{ch}} ( \sqrt{s},  p_{\mathrm{T}}^{\mathrm{min}} ) \rangle \),
 as the results of these distributions fit with fourth-degree polynomial function 
 are presented in Table~\ref{tab:average_nch}. 
The \( \chi^2/ndf\) is good for all fits. 
The average multiplicity as the results of the sum of these distributions are agreed with 
fit results with accuracy in the third digit after coma. 
Therefore the systematic uncertainties for average multiplicities calculated using these methods 
is negligible.

\begin{figure*}[t!]
\begin{minipage}[h]{0.50\textwidth} 
\center{\includegraphics[width=1.0\linewidth]{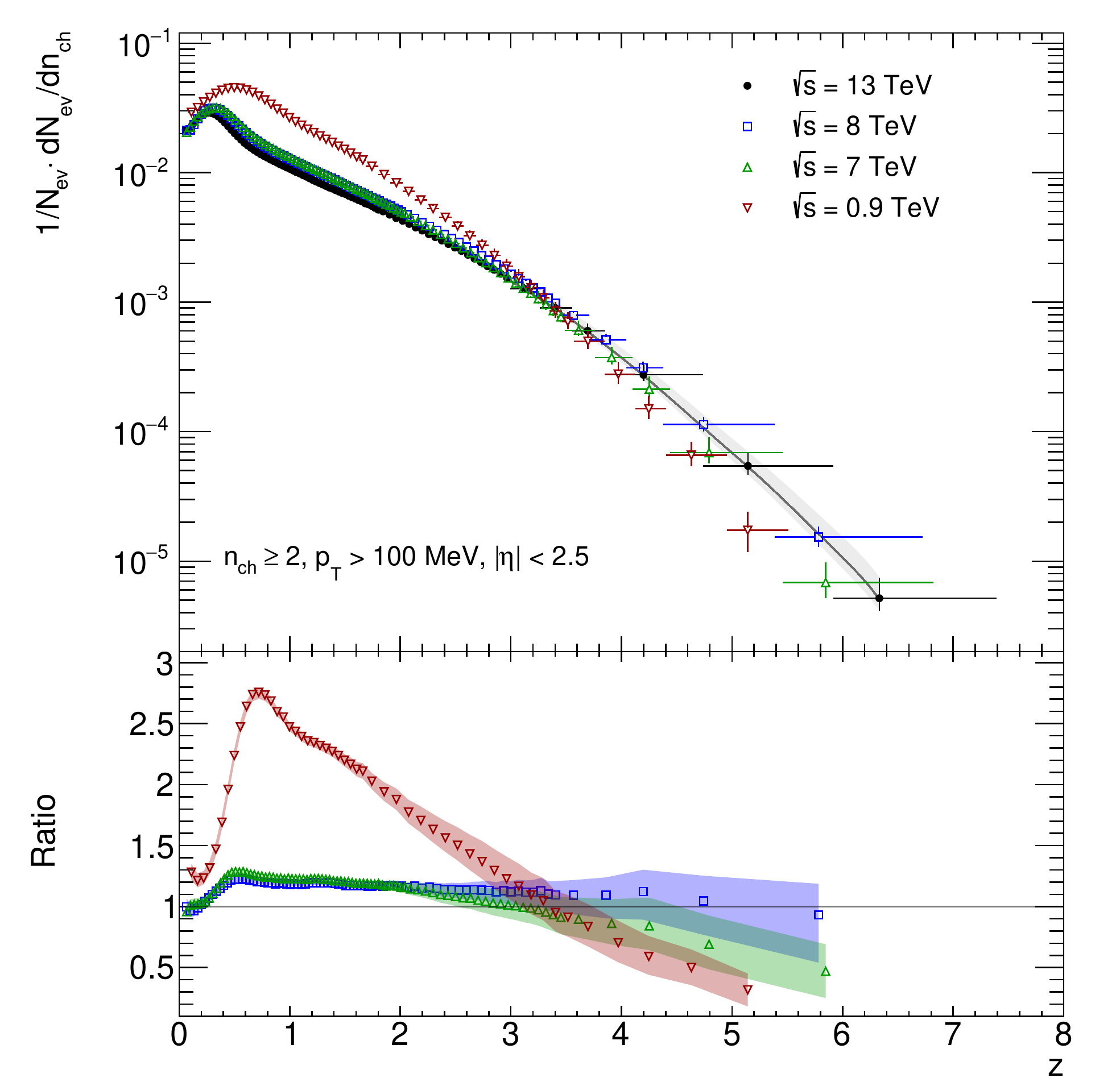}} 
(a) 
\\
\end{minipage}
\hfill
\begin{minipage}[h]{0.50\textwidth} 
\center{\includegraphics[width=1.0\linewidth]{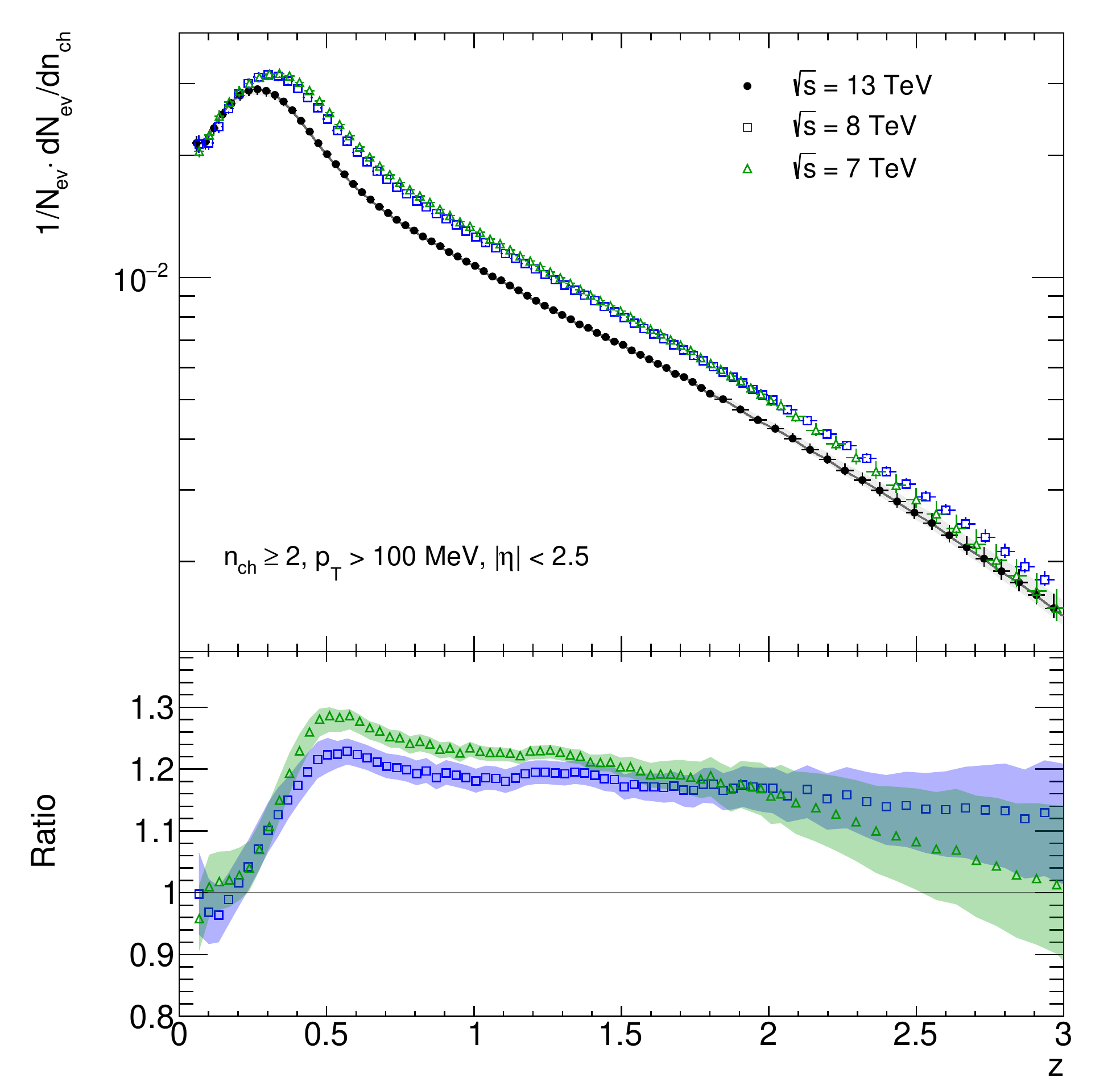}} 
(b)
\\
\end{minipage}
\caption{
Top panel: 
Primary charged-particle  multiplicity distributions    as a function of the scaled multiplicity 
\(z\), defined in Eq.~(\ref{eq_mch}),
for events with  
\(n_{\mathrm{ch}} \ge 2\), \(p_{\mathrm{T}} >100\)~MeV 
and \(\mid\eta\mid  < 2.5\) 
measurement at the centre-of-mass energies 
\(0.9\),  \(7\),  \(8\) and \(13\)~TeV 
by the ATLAS 
\cite{STDM-2010-01,STDM-2010-06,STDM-2014-19,STDM-2015-02,STDM-2015-17}
in 
(a) complete multiplicity region and 
(b) zoom multiplicity region with \(z \le 3\) 
at the \( \sqrt{s} = 7\),  \(8\) and \(13\)~TeV.
%
The gray curve and band of the uncertainties are the result of the interpolation 
of the charged-particle multiplicity distribution at  \(13\)~TeV.
The error bars and boxes represent the statistical and systematic contributions, respectively.
Bottom panel: 
The ratios of the charged-particle  multiplicity distributions to the interpolated distribution at \( \sqrt{s}  = 13\)~TeV are shown. 
%
%
Bands represent the uncertainties for the ratios 
as results of statistical and systematic uncertainties added in quadrature for both distributions.
%
%
}
\label{fig_events_pT100_mch}
\end{figure*}

\begin{figure*}[t!]
\begin{minipage}[h]{0.50\textwidth} 
\center{\includegraphics[width=1.0\linewidth]{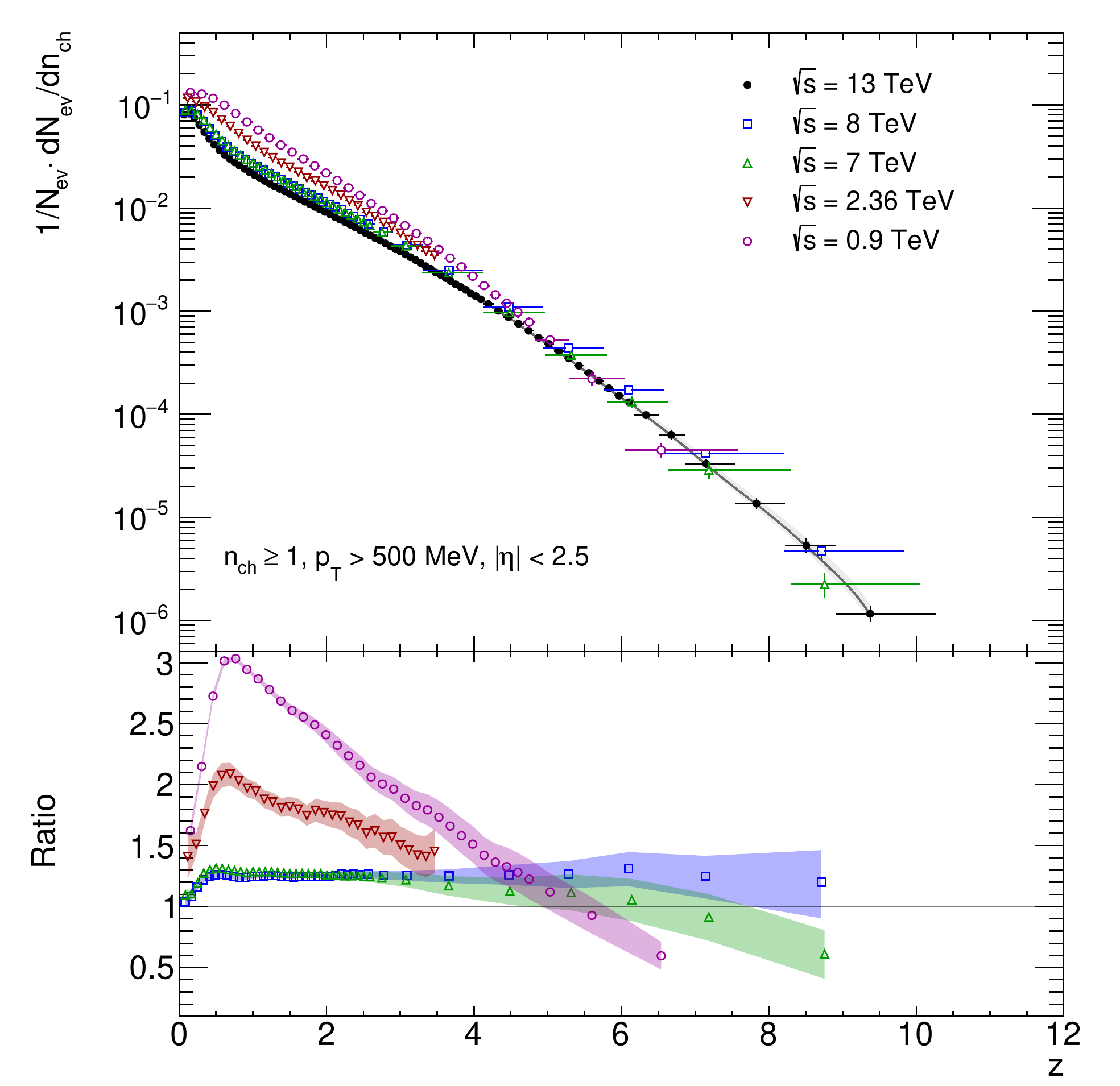}} 
(a) 
\\
\end{minipage}
\hfill
\begin{minipage}[h]{0.50\textwidth} 
\center{\includegraphics[width=1.0\linewidth]{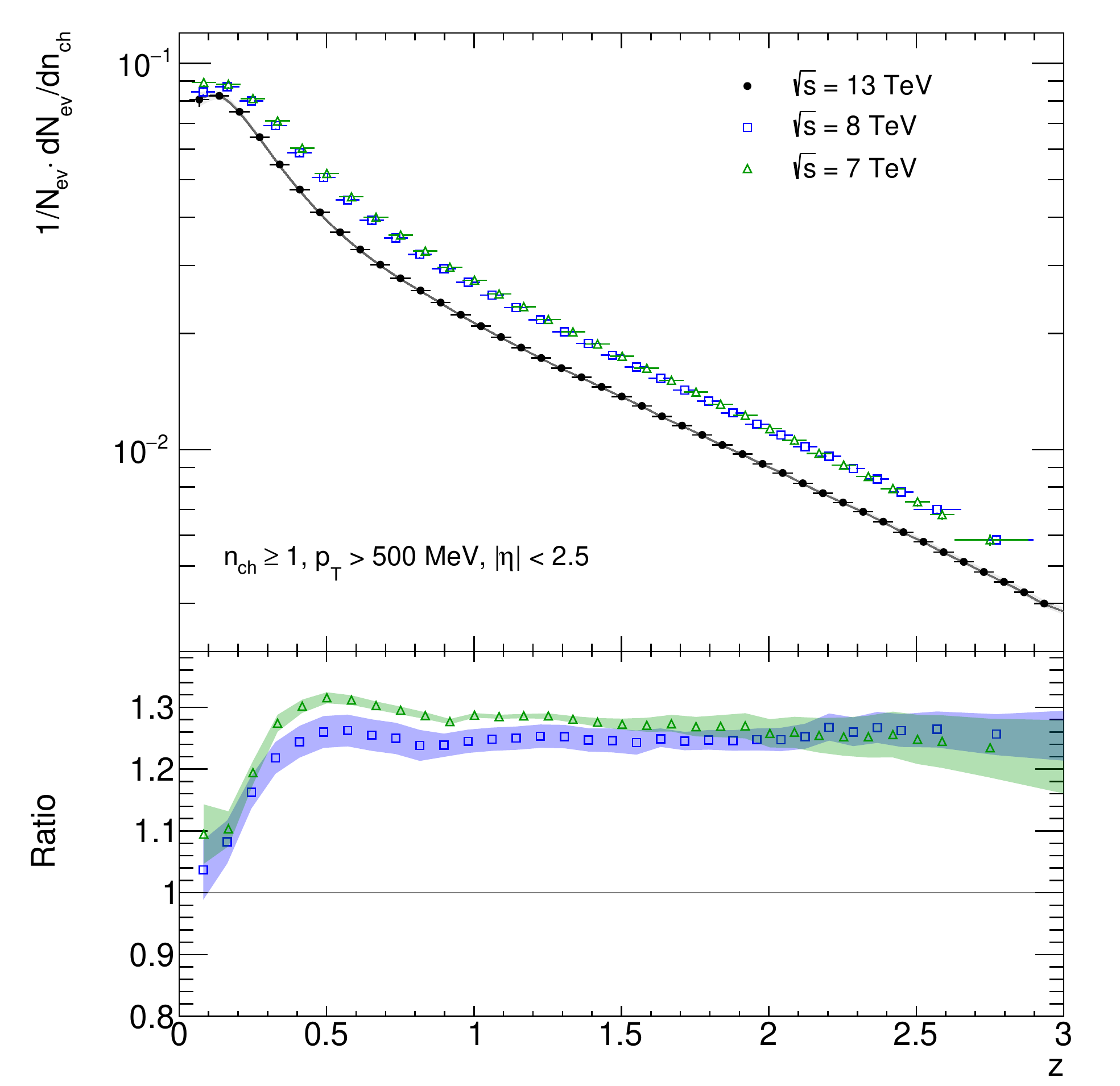}} 
(b)
\\
\end{minipage}
\caption{
Top panel: 
Primary charged-particle  multiplicity distributions  as a function of the scaled multiplicity 
\(z\), defined in Eq.~(\ref{eq_mch}),
for events with  \(n_{\mathrm{ch}} \ge 1\), \(p_{\mathrm{T}} >500\)~MeV and \(\mid\eta\mid  < 2.5\) 
measurement  at the centre-of-mass energies  \(0.9\),  \(2.36\), \(7\),  \(8\) and \(13\)~TeV  by the ATLAS 
\cite{STDM-2010-01,STDM-2010-06,STDM-2014-19,STDM-2015-02,STDM-2015-17} in
(a) complete multiplicity region and 
(b) zoom multiplicity region with \(z \le 3\) 
at   the \( \sqrt{s} = 7\),  \(8\) and \(13\)~TeV.
The gray curve and band of the uncertainties are the result of the interpolation 
of the charged-particle multiplicity distribution at  \(13\)~TeV.
The error bars and boxes represent the statistical and systematic contributions, respectively.
Bottom panel: 
The ratios of the charged-particle  multiplicity distributions to the interpolated distribution at \( \sqrt{s}  = 13\)~TeV are shown. 
%
%
Bands represent the uncertainties for the ratios 
as results of statistical and systematic uncertainties added in quadrature for both distributions.
%
}
\label{fig_events_pT500_mch}
\end{figure*}

The comparison of the primary charged-particle multiplicities as a function of 
the scaled multiplicity
\(z\) or KNO scale, defined in Eq.~(\ref{eq_mch}), 
for events with  
\(n_{\mathrm{ch}} \ge 2\) and \(p_{\mathrm{T}} >100\)~MeV;
\(n_{\mathrm{ch}} \ge 1\) and \(p_{\mathrm{T}} >500\)~MeV 
for  \(\mid\eta\mid  < 2.5\) measurement by ATLAS Collaboration  
at  the \( \sqrt{s} \) from \(0.9\)  to  \(13\)~TeV 
\cite{STDM-2010-01,STDM-2010-06,STDM-2014-19,STDM-2015-02,STDM-2015-17}
are presented 
in
Fig.~\ref{fig_events_pT100_mch} 
and  
Fig.~\ref{fig_events_pT500_mch}, respectively. 
For these figures the multiplicity axis was compressed  by the factor 
\( \langle n_{\mathrm{ch}} ( \sqrt{s},  p_{\mathrm{T}}^{\mathrm{min}} ) \rangle \).
The KNO scale is the same and therefore it is correct scale for comparison
distributions for different center-of-mass energies
or distributions for different kinematic regions.

The scaled multiplicity regions are up to  \(7.5\) average multiplicity  for \(p_{\mathrm{T}} >100\)~MeV  
and up to \(10.5\) average multiplicity  for \(p_{\mathrm{T}} >500\)~MeV as shown 
in
Figs.~\ref{fig_events_pT100_mch}(a) and \ref{fig_events_pT500_mch}(a), respectively. 

In Table~\ref{tab:average_nch} shown the relative uncertainty, 
\( \delta\langle n_{\mathrm{ch}} \rangle / \langle n_{\mathrm{ch}}\rangle \), 
for average multiplicities.
Relative uncertainties are small and equal to 
0.32--0.66\% for  \(p_{\mathrm{T}} >100\)~MeV and
0.24--0.46\% for  \(p_{\mathrm{T}} >500\)~MeV, 
except result at \(\sqrt{s}=2.36\)~GeV which was measured with the worst accuracy.
The relative uncertainty for a bin size,  \( \delta z/  \Delta z \),
of KNO variable \(z\) 
is equal to 
\( \delta\langle n_{\mathrm{ch}} \rangle /\langle n_{\mathrm{ch}}\rangle \).
Therefore influence of average multiplicity uncertainties 
presented in Table~\ref{tab:average_nch} 
on KNO scale are very small.

In the bottom panels ratios of charged-particle distributions at \(0.9\) -- \(8\)~TeV 
to the distribution at \( 13\)~TeV are shown. 
Ratios, and their uncertainties,  of charged-particle distributions at smaller centre-of-mass 
energies to the distribution at \(13\)~TeV,  which was obtained by interpolation,  are presented in
Figs.~\ref{fig_events_pT100_mch}, \ref{fig_events_pT500_mch} and following 
Figures in Sec.~\ref{KNO_scaling} and Sec.~\ref{average_pT_z}.
The points in the ratios were obtained as a ratio at smaller centre-of-mass 
energy to the result of the interpolation at \(13\)~TeV because 
KNO scale (\(z\)) and a bin size depend from 
energy.  
For the interpolation procedure the  \textsc{Interpolator}  method  
of the \textsc{Root} statistical analysis framework \cite{Antcheva:2009zz} was used. 
Figures~\ref{fig_events_pT100_mch} -- \ref{fig_averagepT_pT100_mch} 
show the gray curve and band of the uncertainties  as the result of the interpolation of the 
distribution  at  \(13\)~TeV.

Figures~\ref{fig_events_pT100_mch} and \ref{fig_events_pT500_mch}
show that primary charged-particle multiplicity distributions decrease 
as collision energy  increases from \(0.9\)  to  \(13\)~TeV 
on the factor of  \(\approx 3\)  for maximum at \( z \approx 0.7 \).

The results for the \(\sqrt{s} = 7\), \(8\) and \(13\)~TeV and \(z \le 3\) 
are presented in  Fig.~\ref{fig_events_pT100_mch}(b)  for \(p_{\mathrm{T}} >100\)~MeV  
and Fig.~\ref{fig_events_pT500_mch}(b)  for \(p_{\mathrm{T}} >500\)~MeV. 
One can see that for  the distributions at \( \sqrt{s} = 7\) and \(8\)~TeV 
there is  agreement within error bars except for region  \( 0.5 < z < 1.5 \). 
The multiplicity distribution at  \(8\)~TeV is  \(\approx  20\)\% larger 
than at \(13\)~TeV for  region the \( z < 3\) in both cases.

\section{Study of the KNO scaling}
\label{KNO_scaling}

\begin{figure*}[t!]
\begin{minipage}[h]{0.50\textwidth} 
\center{\includegraphics[width=1.0\linewidth]{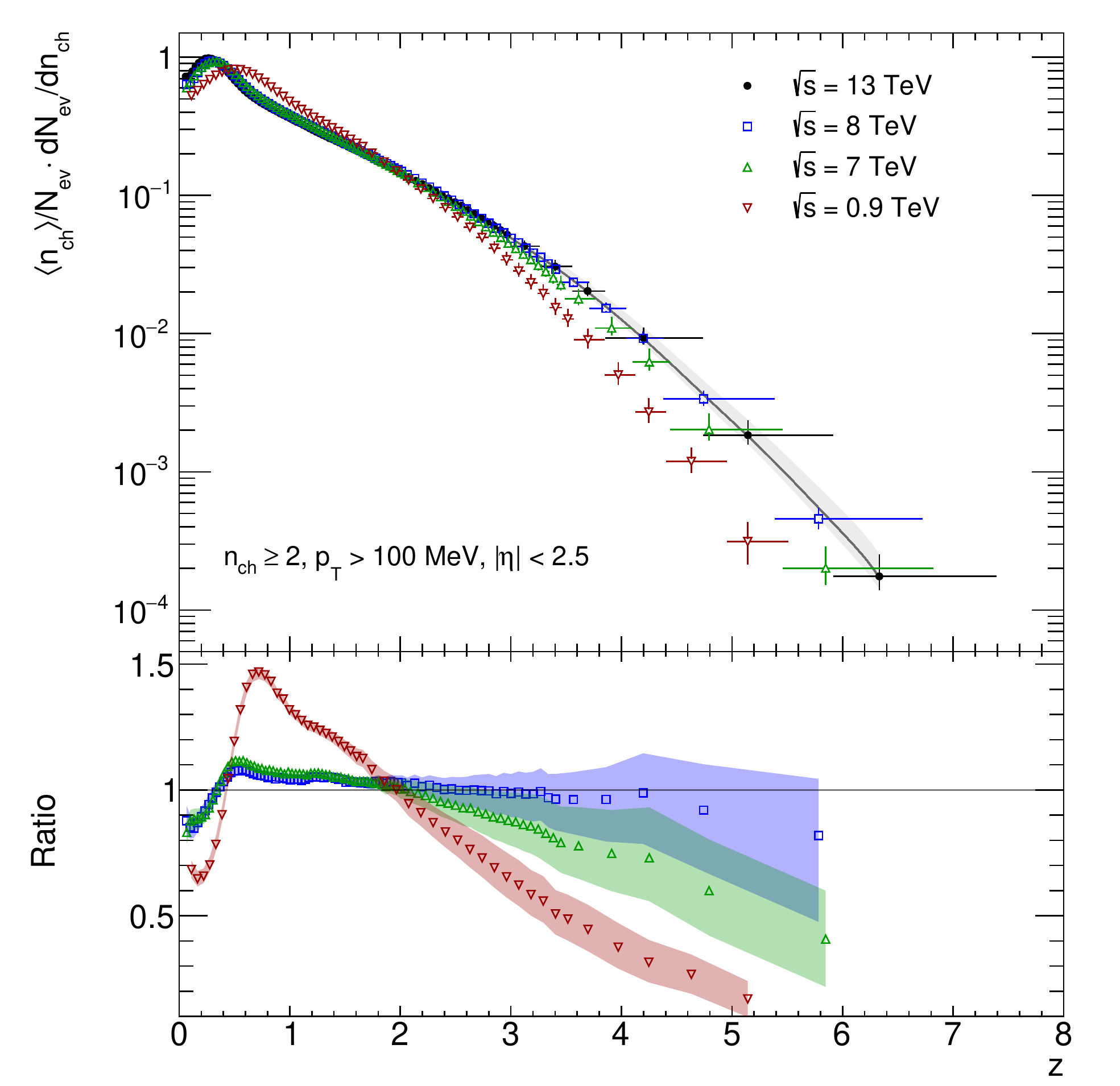}} 
(a) 
\\
\end{minipage}
\hfill
\begin{minipage}[h]{0.50\textwidth} 
\center{\includegraphics[width=1.0\linewidth]{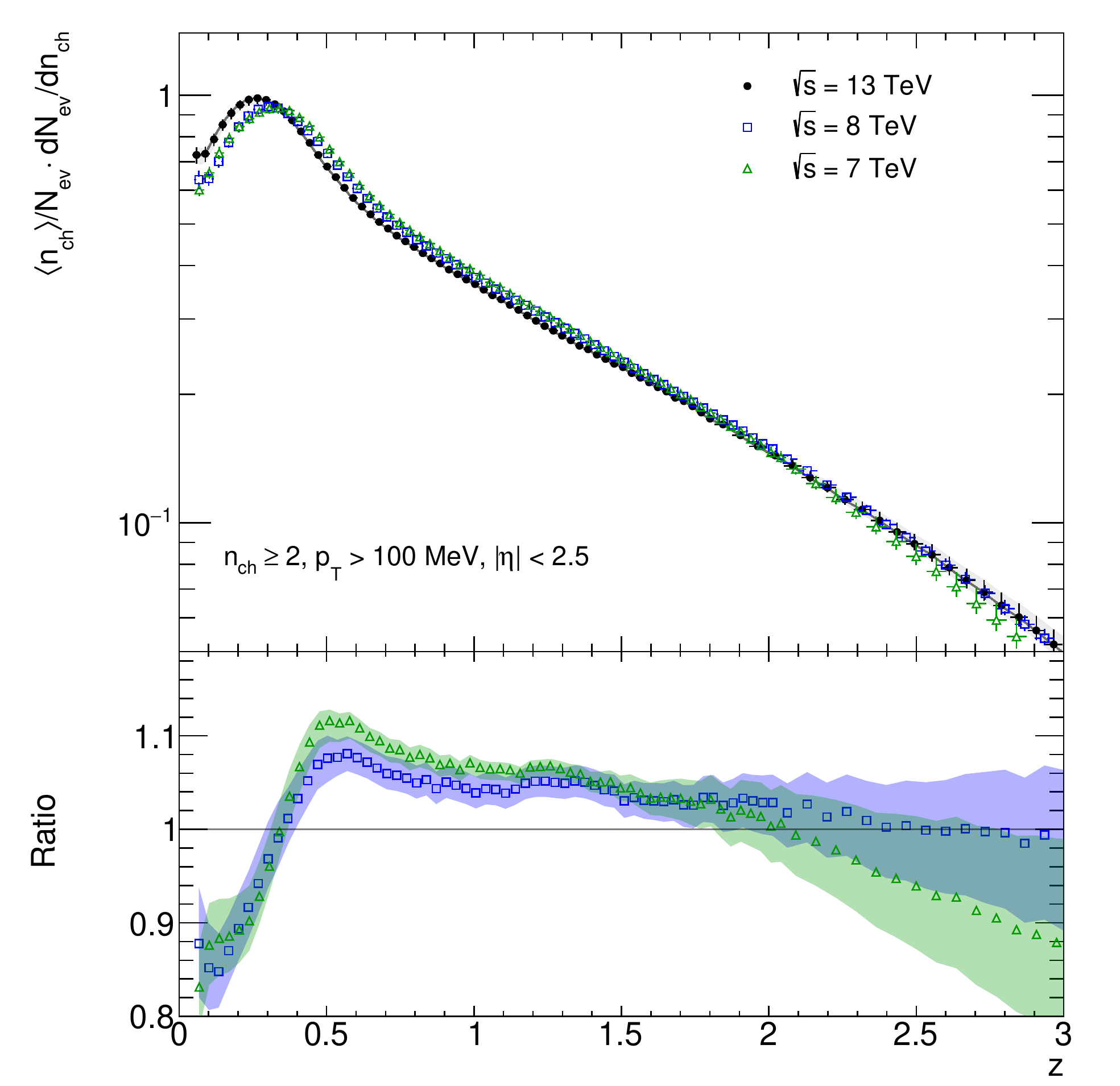}} 
(b)
\\
\end{minipage}
\caption{
Top panel: 
KNO scaled primary charged-particle  multiplicity distributions   
as a function of the scaled multiplicity 
\(z\), defined in Eq.~(\ref{eq_mch}),
for events with  \(n_{\mathrm{ch}} \ge 2\), \(p_{\mathrm{T}} >100\)~MeV and \(\mid\eta\mid  < 2.5\) 
measurement  at the centre-of-mass energies 
\(0.9\),  \(7\),  \(8\) and \(13\)~TeV 
by the ATLAS 
\cite{STDM-2010-01,STDM-2010-06,STDM-2014-19,STDM-2015-02,STDM-2015-17}
in 
(a) complete multiplicity region and 
(b) zoom multiplicity region with \(z \le 3\) 
at  the  \(\sqrt{s} = 7\),  \(8\) and \(13\)~TeV.
The gray curve and band of the uncertainties are the result of the interpolation 
of the charged-particle multiplicity distribution at  \(13\)~TeV.
The uncertainties represent the sum in quadrature of the statistical and systematic contributions.
Bottom panel: 
The ratios of the KNO scaled primary charged-particle distributions to the interpolated distribution at \( \sqrt{s}  = 13\)~TeV are shown. 
%
%
Bands represent the uncertainties for the ratios 
as results of statistical and systematic uncertainties added in quadrature for both distributions.
%
}
\label{fig_events_pT100_mch_KNO}
\end{figure*}

\begin{figure*}[t!]
\begin{minipage}[h]{0.50\textwidth} 
\center{\includegraphics[width=1.0\linewidth]{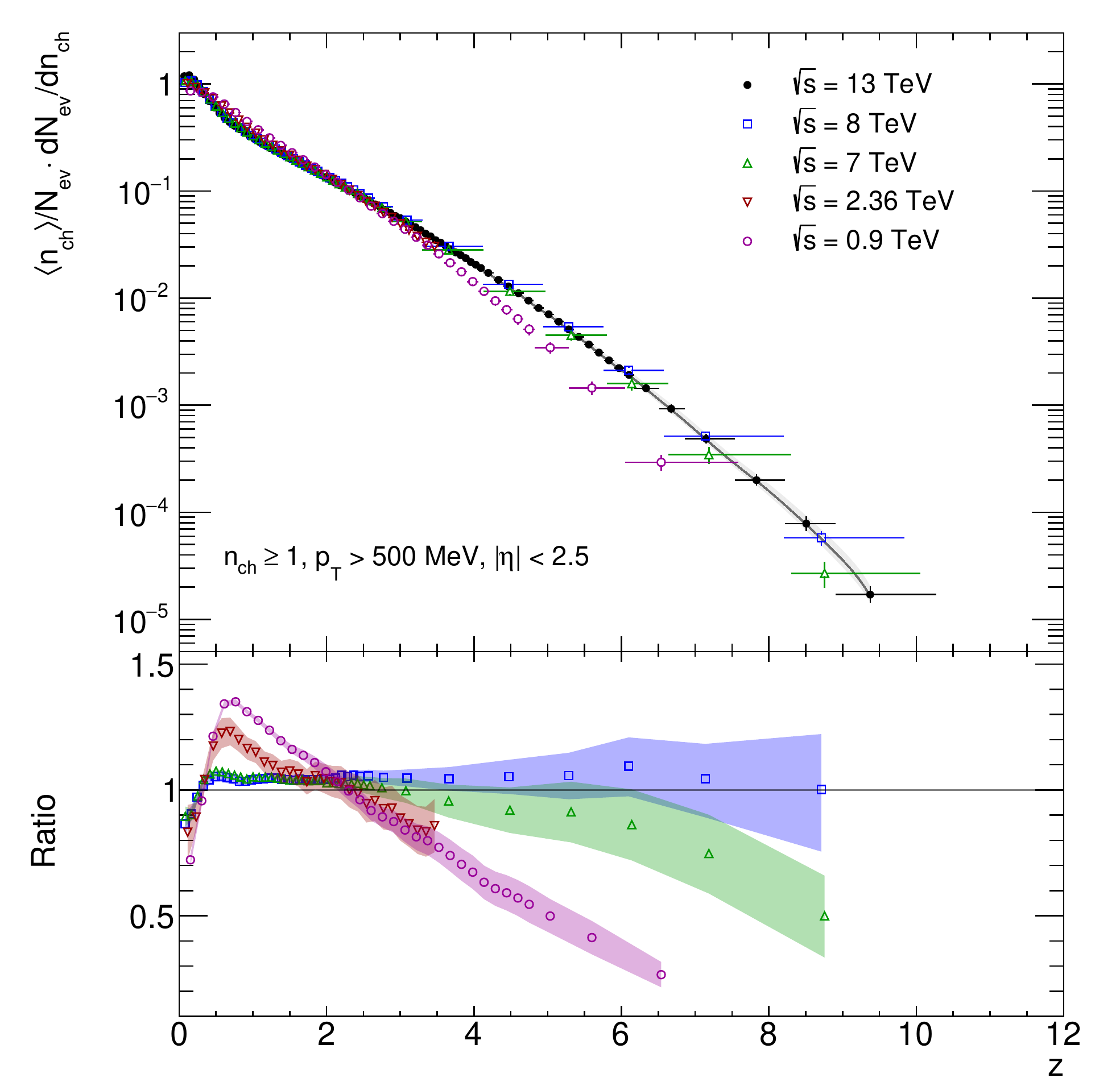}} 
(a) 
\\
\end{minipage}
\hfill
\begin{minipage}[h]{0.50\textwidth} 
\center{\includegraphics[width=1.0\linewidth]{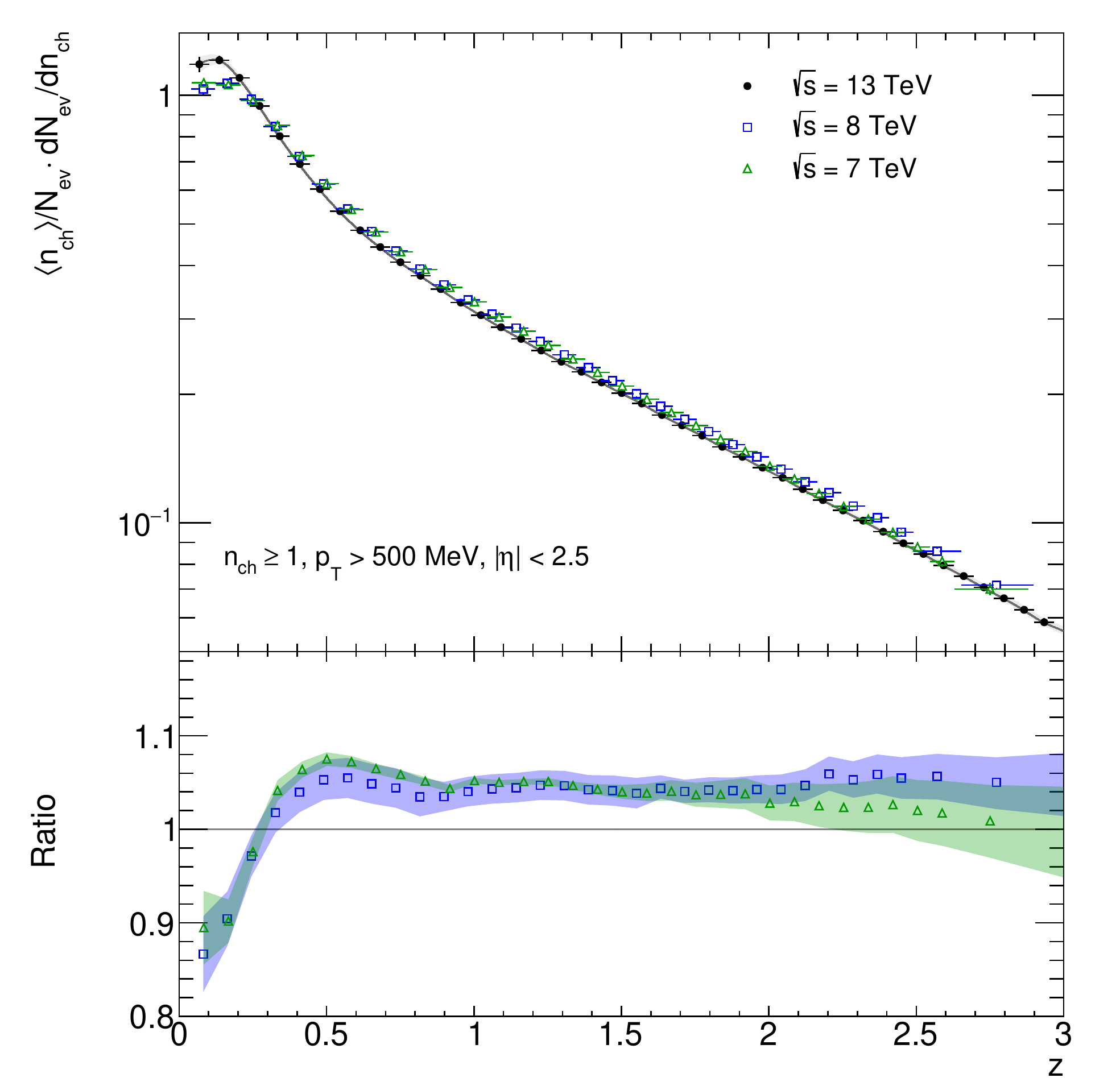}} 
(b)
\\
\end{minipage}
\caption{
Top panel: 
KNO scaled primary charged-particle multiplicity distributions  as a function of the scaled multiplicity 
\(z\), defined in Eq.~(\ref{eq_mch}),
for events with  \(n_{\mathrm{ch}} \ge 1\), \(p_{\mathrm{T}} >500\)~MeV and \(\mid\eta\mid  < 2.5\) 
measurement  at the centre-of-mass energies 
\(0.9\),  \(2.36\), \(7\),  \(8\) and \(13\)~TeV 
by the ATLAS 
\cite{STDM-2010-01,STDM-2010-06,STDM-2014-19,STDM-2015-02,STDM-2015-17}
in 
(a) complete multiplicity region and 
(b) zoom multiplicity region with \(z \le 3\) 
at  the  \(\sqrt{s} = 7\),  \(8\) and \(13\)~TeV.
The gray curve and band of the uncertainties are the result of the interpolation 
of the charged-particle multiplicity distribution at  \(13\)~TeV.
The uncertainties represent the sum in quadrature of the statistical and systematic contributions.
Bottom panel: 
The ratios of the KNO scaled primary charged-particle distributions to the interpolated distribution at \( \sqrt{s}  = 13\)~TeV are shown. 
%
%
Bands represent the uncertainties for the ratios 
as results of statistical and systematic uncertainties added in quadrature for both distributions.
%
}
\label{fig_events_pT500_mch_KNO}
\end{figure*}

The KNO scale variable   \(z \) in Eq.~(\ref{eq_mch})  
provides  the way to study the evolution of the shape of KNO charged-particle multiplicity distributions 
(\ref{eq_Psi_nch})
with varying centre-of-mass energy   and kinematic region, for example  \(p_{\mathrm{T}}\) threshold.
The  KNO distributions (\ref{eq_Psi_nch}) and their ratios are presented in
Fig.~\ref{fig_events_pT100_mch_KNO} 
for charged particles with \(p_{\mathrm{T}} >100\)~MeV
and
Fig.~\ref{fig_events_pT500_mch_KNO} 
for charged particles with \(p_{\mathrm{T}} >500\)~MeV.
These figures are similar to  Fig.~\ref{fig_events_pT100_mch} and Fig.~\ref{fig_events_pT500_mch} 
but 
the vertical axis is stretched by
the factor 
\( \langle n_{\mathrm{ch}} (\sqrt{s}, p_{\mathrm{T}}^{\mathrm{min}})\rangle \).

The quantities of interest are derived from the original set of  nine KNO distributions and the 
ratios 
of  these distributions to the one at \(13\)~TeV. 
The high-multiplicity tail of distributions is pushed up and the maximum of the distribution 
is shifted towards small values of \(z\) with collision energy increase.

Ratios of the KNO distributions between the smallest centre-of-mass energy \(0.9\) and  \(13\)~TeV 
exceed the maximum positive value at  \( z \approx 0.8 \) 
and the maximum negative value for the highest multiplicity  at \( z \approx 5.5 \) 
for \(p_{\mathrm{T}} >100\)~MeV 
(Fig.~\ref{fig_events_pT100_mch_KNO}(a))
and
\( z \approx 6.5 \) for \(p_{\mathrm{T}} >500\)~MeV 
(Fig.~\ref{fig_events_pT500_mch_KNO}(a)).
There is an intersection point for all distributions at \( z \approx 2\).

A test of the KNO scaling distributions between \(\sqrt{s} = 0.9\) and \(13\)~TeV 
confirms that KNO scaling violation increases with decreasing collision energy.
Ratios of the KNO distributions  between the highest energies \(8\)  and \(13\)~TeV exceed   
the maximum value of 
\(+8\)\% at  \( z \approx 0.5 \)  and   the minimum value of  \(-15\)\% at \( z \approx 0.1 \)   for \(p_{\mathrm{T}} >100\)~MeV 
(Fig.~\ref{fig_events_pT100_mch_KNO}(b))
and  the maximum value of  \(+5\)\% at  \( z \approx 0.5 \)  and  \( -13 \)\% at \( z \approx 0.1 \)  for \(p_{\mathrm{T}} >500\)~MeV 
(Fig.~\ref{fig_events_pT500_mch_KNO}(b)).
For the high multiplicity tail, these ratios are in agreement within error bars with the KNO distribution at \(13\)~TeV.

Single- and double-diffractive processes give an important contribution only for the low multiplies region, \(z \lesssim 0.3\).
The typologies of diffractive and non-diffractive events are different and their KNO behavior may be also different. 
The negative spread, \( \lesssim -8 \)\%, for the low multiplicity may be the results of diffractive processes 
contributing.

The KNO scaling  
tends 
to be independent of energy  at \(\sqrt{s} = 8\) and \(13\)~TeV
within   \( \approx \mbox{}^{\mbox{~}+8}_{-15} \)\% for \( z \lesssim 2 \) and  within error bars for \( z \gtrsim 2  \)  
for events with charged-particle transverse momentum  \(p_{\mathrm{T}} >100\)~MeV  
(Fig.~\ref{fig_events_pT100_mch_KNO}(b)), 
and  within   \( \mbox{}^{\mbox{~}+5}_{-13} \)\% for \( z \lesssim 3 \) and  within error bars for \( z \gtrsim 3  \)  
for events with charged-particle transverse momentum  \(p_{\mathrm{T}} >500\)~MeV 
(Fig.~\ref{fig_events_pT500_mch_KNO}(b)). 
The tendency of the KNO scaling to 
hold 
for the highest collision energies is observed.  

The MC QGSM predictions are made for the KNO non-diffractive charged-particle multiplicity distributions 
for \(pp\) collisions including at highest LHC centre-of-mass energy \(\sqrt{s}= 14\)~TeV 
for \(\mid\eta\mid <2.4\) on Fig.~12 in Ref.~\cite{Bleibel:2010ar}.
These distributions are the same qualitative behaviour as those presented in 
Fig.~\ref{fig_events_pT100_mch_KNO}(a).
The MC QGSM described the KNO  distributions as the contribution of
the cylinder diagram and diagrams with multi-Pomeron scattering. 
%
The pronounced peak in the low \(z\)  arises solely due to a single Pomeron exchange
and the maxima of distributions for multi-Pomeron processes are moved in the direction of 
high \(z\)  thus pushed up the tail \cite{Bleibel:2010ar}.

\section{Average transverse momentum dependences}
\label{average_pT_z}

\begin{figure*}[t!]
\begin{minipage}[h]{0.50\textwidth}
\center{\includegraphics[width=1.0\linewidth]{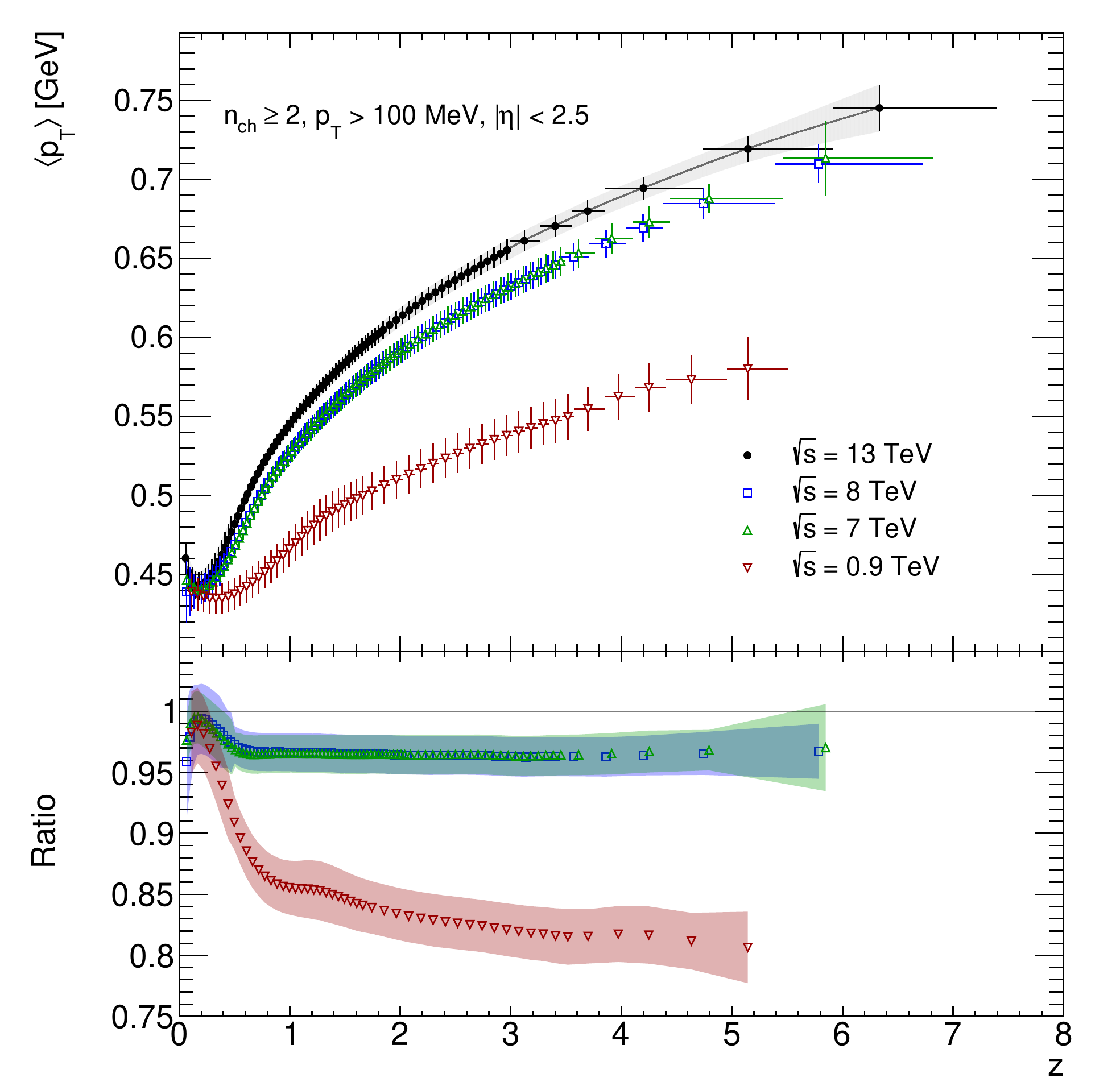}} 
(a)
\\
\end{minipage}
\hfill
\begin{minipage}[h]{0.50\textwidth}
\center{\includegraphics[width=1.0\linewidth]{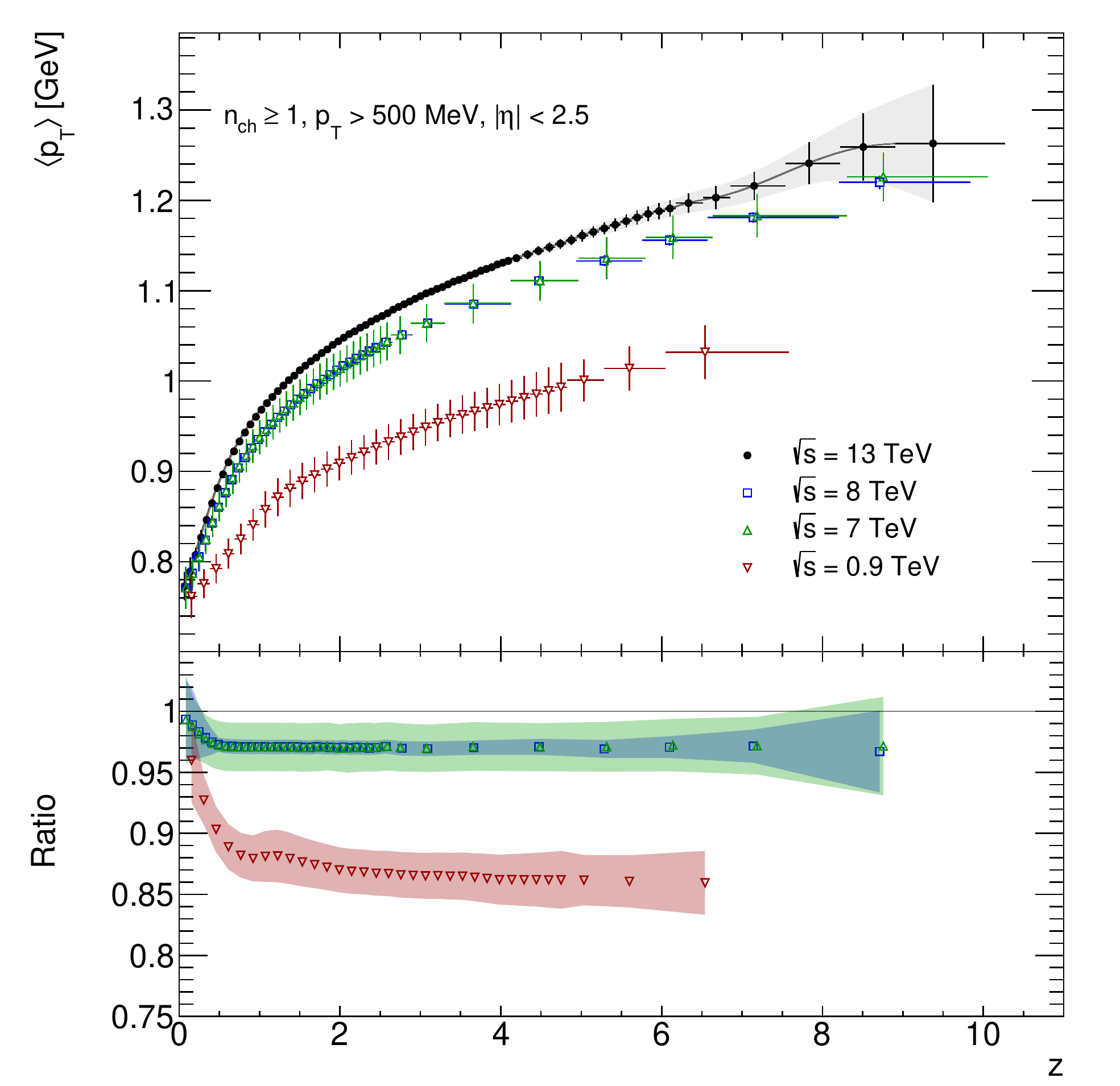}} 
(b)
\\
\end{minipage}
\caption{
Top panel: 
The average transverse momentum, 
\(\langle p_{\mathrm{T}}\rangle \),
as a function of the scaled multiplicity 
\(z\), defined in Eq.~(\ref{eq_mch}),
for events with 
(a)  \(n_{\mathrm{ch}} \ge 2\),   \(p_{\mathrm{T}} >100\)~MeV 
and 
(b)  \(n_{\mathrm{ch}} \ge 1\),   \(p_{\mathrm{T}} >500\)~MeV 
for  \(\mid\eta\mid  < 2.5\) 
measurement   at the  centre-of-mass energies 
\(0.9\), \(7\), \(8\) and  \(13\)~TeV 
by the ATLAS 
\cite{STDM-2010-01,STDM-2010-06,STDM-2014-19,STDM-2015-02,STDM-2015-17}. 
(a) The gray curve and band of the uncertainties are the result of the interpolation 
of the charged-particle multiplicity distribution at 
\(13\)~TeV.
The gray curve and band of the uncertainties are the result of the interpolation 
of the charged-particle multiplicity distribution at  \(13\)~TeV.
The error bars and boxes represent the statistical and systematic contributions, respectively.
Bottom panel: 
The ratios of the average transverse momentum distributions to the interpolated distribution at \( \sqrt{s}  = 13\)~TeV are shown. 
%
%
Bands represent the uncertainties for the ratios 
as results of statistical and systematic uncertainties added in quadrature for both distributions.
%
}
\label{fig_averagepT_pT100_mch}
\end{figure*}

The correct comparison 
of the primary charged-particle average transverse momentum, 
\(\langle p_{\mathrm{T}}\rangle \), as a function of the  scaled multiplicity 
\( z\) (\ref{eq_mch})  
for events with  
\(n_{\mathrm{ch}} \ge 2\) and \(p_{\mathrm{T}} >100\)~MeV;
\(n_{\mathrm{ch}} \ge 1\) and \(p_{\mathrm{T}} >500\)~MeV 
for  \(\mid\eta\mid  < 2.5\) measurement 
at the  centre-of-mass energies from \(0.9\)  to  \(13\)~TeV 
by the ATLAS 
\cite{STDM-2010-01,STDM-2010-06,STDM-2014-19,STDM-2015-02,STDM-2015-17}
are presented in 
Fig.~\ref{fig_averagepT_pT100_mch}. 

Figures~\ref{fig_averagepT_pT100_mch}(a) and (b)
show  an increase of the average transverse momentum distributions 
with the scaled multiplicity.
The \(\langle p_{\mathrm{T}}\rangle \) distribution as a function of \( z \) acquires higher value at higher collision energies.
The value of \(\langle p_{\mathrm{T}}\rangle \) increases by \( 18\)\%  and  \(13\)\%  for \(z > 1\)  
with energy increase from \(0.9\) to \(13\)~TeV  for 
\(p_{\mathrm{T}} >100\)~MeV and \(p_{\mathrm{T}} >500\)~MeV, 
respectively. 
The results at \(7\) and \(8\)~TeV are in agreement within error bars. 
The value of \(\langle p_{\mathrm{T}}\rangle \) increases by 
\(\approx 3\)\% for \(p_{\mathrm{T}} >100\)~MeV 
and  by \(\approx 2.5\)\% for \(p_{\mathrm{T}} >500\)~MeV
with increase in energy from \(8\) to \(13\)~TeV  for  \(z > 0.5\).
The ratio of  \(\langle p_{\mathrm{T}}\rangle \) for  \(8\) to \(13\)~TeV 
are in \(\approx 6\) times  smaller than the ratio for \(0.9\)  to  \(13\)~TeV.

\section{Conclusion}

The comparisons of the charged-particle multiplicity   and  the average transverse momentum distributions 
on the scaled multiplicity, KNO scale, using the results of the ATLAS collaboration at the LHC were presented.
These distributions were measured in proton-proton collisions at centre-of-mass energies  
 \(\sqrt{s} = 0.9\), \( 2.36\), \(7\), \(8\) and \(13\)~TeV   
for the absolute pseudorapidity region less than \(2.5\) and 
and  for two events samples
\(n_{\mathrm{ch}} \ge 2\),  \(p_{\mathrm{T}} > 100\)~MeV
and  
\(n_{\mathrm{ch}} \ge 1\),  \(p_{\mathrm{T}} > 500\)~MeV.
The charged-particle multiplicity distributions on the KNO scale 
have the  similar shape  and decrease with increasing energy. 
The study of the KNO scaling using the ATLAS  results was done.
A test of the KNO scaling between \(0.9\)  and \(13\)~TeV  
confirms that the KNO scaling violation increases with decreasing collision energy.
The KNO distributions tend to be independent of energy for the highest energies.
The mean transverse momentum on the KNO scale has the same shape and increases with increasing energy. 

\section*{Acknowledgements}
We thank the ATLAS collaboration for the excellent experimental results which were used for this analysis.
Special thanks to 
E.~K.~Sarkisyan-Grinbaum 
and 
S.~Ya.~Tokar 
for several productive discussions. 
Thanks  
to E.~E.~Zabrodin 
for fruitful discussions concerning 
QGSM predictions.  
Useful discussions with  
V.~V.~Glagolev, 
G.~I.~Lykasov
and
N.~A.~Russakovich
are gratefully acknowledged.

\printbibliography

\end{document}